\documentclass[aps,prd,groupedaddress,showpacs,showkeys]{revtex4}
\usepackage{epsfig}
\usepackage{graphicx}
\usepackage{amsmath}
\usepackage{amssymb}
\usepackage{amsfonts,tabularx}

\begin{document}

\title{$\phi$ meson production in $p\overline p$ annihilation}
\author{S. Srisuphaphon\footnotemark[1],
  Y. Yan \footnotemark[1]\footnotemark[3],
  Thomas Gutsche\footnotemark[2],
  Valery E. Lyubovitskij\footnotemark[2]\footnotemark[4]
}
\affiliation{ \footnotemark[1] School of Physics, Suranaree
University of Technology, 111 University Avenue, Nakhon Ratchasima
30000, Thailand
\\
 \footnotemark[2]
Institut f\"ur Theoretische Physik,
Universit\"at T\"ubingen, \\
Kepler Center for Astro and Particle Physics,
\\ Auf der Morgenstelle 14, D-72076 T\"ubingen, Germany
\\
\footnotemark[3]
Thailand Center of Excellence in Physics, Ministry
of Education, Bangkok, Thailand \email{yupeng@sut.ac.th}
\\
\footnotemark[4]
On leave of absence
from Department of Physics, Tomsk State University,
634050 Tomsk, Russia
\vspace*{0.4\baselineskip}}

\date{\today}

\begin{abstract}
Apparent channel-dependent violations of the OZI rule in
nucleon-antinucleon annihilation reactions are discussed in the
presence of an intrinsic strangeness component in the nucleon.
Admixture of $s\bar s$ quark pairs in the nucleon wave function
enables the direct coupling to the $\phi$ - meson in the
annihilation channel without violating the OZI rule. Three forms are
considered in this work for the strangeness content of the proton
wave function, namely, the $uud$ cluster with a $s\bar{s}$ sea quark
component, kaon-hyperon clusters based on a simple chiral quark
model, and the pentaquark picture $uuds\bar s$. Nonrelativistic
quark model calculations reveal that the strangeness magnetic moment
$\mu_s$ and the strangeness contribution to the proton spin
$\sigma_s$ from the first two models are consistent with recent
experimental data where $\mu_s $ and $\sigma_s $ are negative. For
the third model, the $uuds$ subsystem with the configurations
$[31]_{FS}[211]_{F}[22]_{S}$ and $[31]_{FS}[31]_{F}[22]_{S} $ leads
to negative values of $\mu_s $ and $\sigma_s $. With effective quark
line diagrams incorporating the $^3P_0$ model we give estimates for
the branching ratios of the annihilation reactions at rest $p
\bar{p} \rightarrow \phi X$ ($X=\pi^0,\eta,\rho^0,\omega$). Results
for the branching ratios of $\phi X$ production from atomic $p\bar
p$ s-wave states are for the first and third model found to be
strongly channel dependent, in good agreement with measured rates.

\end{abstract}



\maketitle

\newpage

\section{Introduction}

In the simple constituent quark model, where the proton is made of
two constituent $u$ quarks and one $d$ quark, a good explanation of
static properties e.g. magnetic moment can be achieved. However,
experimental results of the pion-nucleon sigma term value, strange
magnetic moment $\mu_s$, strangeness contribution to nucleon form
factor \cite{vonHarrach:2005at} as well as the apparent violations
in nucleon-antinucleon annihilation reactions involving $\phi$ meson
\cite{Amsler:1991hs} indicate that the proton might contain a
substantial strange quark-antiquark ($s\bar{s}$) component. The
strangeness sigma term appears to lie somewhere in the range of
$2-7$\% of the nucleon mass \cite{Young:2010aj}. The substantial
Okubo-Zwieg-Iizuka (OZI) rule violations in the $N\bar{N}$
annihilation reactions involving $\phi$ meson may suggest the
presence of an intrinsic $s\bar{s} $ in nucleon wave function
\cite{Ellis:1994ww}, for instance, the presence of a
$q^3s\bar{s}(\bar{q}^3s\bar{s})$ piece in the $N(\bar{N})$ wave
function. With such an assumption, the $\phi $ meson could be
produced in $N\bar{N}$ annihilation reactions via a shake-out or
rearrangement of the strange quarks already stored in the nucleon
without the violation of the OZI rule. There are other explanations
of the OZI rule violation without introducing strange component in
the nucleon such as the resonance interpretation, instanton induced
interaction \cite{Kochelev:1995kc}, and rescattering
\cite{Locher:1994cc}.

The EMC spin experiment \cite{Ashman:1987hv} on deep inelastic
scattering of longitudinally polarised muons by longitudinally
polarised protons revealed the first time that the polarization of
the strange quark sea may contribute to the proton spin $\sigma_s$ a
significant negative value. This experimental result was confirmed by
the subsequent deep inelastic double polarization experiments. Ref.
\cite{Ellis:1994py} analyzed all the available data then in a
systematic way and found $\sigma_s= -0.10 \pm 0.03$. Among a large
number of theoretical works, Cheng and Li apply the chiral quark
model (ChQM) to explain the spin and flavor structure of proton
\cite{Li:1997kp}. With the fluctuation of the proton into a kaon and
a hyperon, they can explain the negative polarization of the strange
quark sea and get other theoretical results consistent with the DIS
experimental results.

However, the configuration of strange quarks in the nucleon is still
an open question.  The strangeness magnetic moment $\mu_s $ can be
extrapolated from the strange magnetic form factor $G_M^s(Q^2)$ at the
momentum transfer $Q^2=0$ measured in the parity violation experiments
of electron scattering from a nucleon \cite{Diehl:2007uc}. Most
experimental measurements suggest a positive value for $\mu_s $, in
contrast to the recent experiment data \cite{Baunack:2009gy} and
most theoretical calculations which have obtained negative values
for this observable~\cite{Beck:2001yx,Lyubovitskij:2002ng}.
A recent work~\cite{An:2005cj}
has proposed a different form for the strangeness
content of the proton which has the strange quark piece in terms of
pentaquark configurations instead of the 5-quark component which
consist of a $uud$ cluster and a $s\bar{s}$ pair proposed for
solving the puzzle of violation the OZI rule. Different pentaquark
configurations that may be contained in the proton may yield both
positive and negative values for the strangeness spin and magnetic
moment of the proton.

The experimental results on $\mu_s$, which is extracted from
experimental data on $G_M^s(Q^2)$, are rather uncertain due to the
large uncertainties in $G_M^s(Q^2)$ and the extrapolation approach. So
it is believed that the proton-antiproton reactions involving $\phi$
production may be another platform to be applied to tackle the
possible configuration of strange quarks in the proton. In the
present work we consider the strange content in the proton wave
function in three models, namely, the $uud$ cluster with a
$s\bar{s}$ sea quark component, kaon-hyperon clusters based on the
chiral quark model, and the pentaquark picture $uuds\bar s$. The
theoretical $\sigma_s$, $\mu_s$ and branching ratios of the
reactions $p\bar{p}\rightarrow \phi X$ ($X
=\pi^0,\eta,\rho^0,\omega$) will be compared to experimental data.
We resort to the $^3P_0$ quark model \cite{LeYaouanc:1988fx} and the
nearest threshold dominance model \cite{Vandermeulen:1988hh} to
obtain quantitative predictions for the branching ratios of the
annihilation reactions from atomic $p\bar{p}$ states with the
relative orbital angular momentum $L=0$ \cite{Gutsche:1997gy}. The
paper is organized as follows. The proton wave functions are briefly
described in Section 2 while $\sigma_s$ and $\mu_s$ are calculated
and discussed in Section 3 for various strangeness quark
configurations. In Section 4 we evaluate the branching ratios for
the reactions $p\bar{p}\rightarrow \phi X$ for the three forms of
proton wave functions by using the $^3P_0$ quark model. Finally a
summary and conclusion are given in Section 5.

\section{Proton wave functions}\label{sec:1}
The proton wave function in the presence of strange quarks may
include a 5-quark component $qqqs\bar{s}$ in addition to the $uud$
quark component, taking generically the form
\begin{equation}
|p\rangle = A|uud\rangle+B|uuds\bar{s}\rangle
\end{equation}
where $A$ and $B$ are the amplitudes for the 3- and 5-quark
components in the proton, respectively \cite{Dover:1990ic}. The
possible spin-flavor structures of the 5-quark components discussed
in the $N\bar N$ annihilation process are considered in
the next three subsections.

\subsection{Proton wave function with an explicit $s\bar{s}$ sea-quark component}
 We consider the idea that strange quarks are present in the form of
 an $s\bar{s}$ sea-quark component in the proton state.
 This idea was proposed for describing the apparent violation
 of the OZI rule in the $\phi NN$ production process~\cite{Henley:1991ge}
 and in more general form used to discuss the  $\phi $ meson production in  $N\bar{N}$
annihilation reactions~\cite{Ellis:1994ww}. The corresponding
5-quark component for this model can be written in Fock space as
\begin{equation}
|uuds\bar{s}\rangle^{s\bar{s}} =  a_0|(uud)_{1/2}(s\bar{s})_0\rangle_{1/2}+a_1|(uud)_{1/2}(s\bar{s})_1\rangle_{1/2}
\end{equation}
where the subscripts denote the spin coupling of the quark clusters,
$a_0$ and $a_1$ represent the amplitudes for the spin 0 and spin 1
components of the admixed $s\bar{s}$ pairs.
\subsection{Proton wave function based on a chiral quark model}
In the chiral quark model, the dominant process is the fluctuation
of a valence quark $q$ into a quark $q'$ plus a Goldstone boson
(GB) which in turn forms a ($q \bar{q}'$) system
\cite{Eichten:1991mt}. After the fluctuation of the $u$ and $d$
quarks in the proton, one of these quarks turns into a quark plus a
quark-antiquark pair involving a strange quark. This idea was
considered, for example, for calculating the flavor and spin content
of the proton \cite{Li:1997kp}. To obtain the proton wave function
we consider the SU(3) invariant interaction Lagrangian
of baryon octet with nonet of pseudoscalar mesons:
\begin{equation}
\mathcal{L}_I=-g_{8}\sqrt{2}\left(\alpha
[\bar B B P]_F +(1-\alpha)[\bar B B P]_D
\right)-g_{1}\frac{1}{\sqrt{3}}[\bar B B P]_S
\end{equation}
where $g_{8}=3.8$ and $g_{1}=2.0$ are coupling constants
\cite{Stoks:1998xv} and $\alpha$ is known as the $F/(F+D)$ ratio
with $F\simeq0.51,~D\simeq0.76$ \cite{Thomas:2001kw}. The square
parentheses denote the SU(3) invariant combinations:
\begin{eqnarray}
{[\bar B B P]}_F&=&
{\rm Tr}(\bar B P B)-{\rm Tr}(\bar B B P)
\,, \\
{[\bar B B M]}_D&=&
{\rm Tr}(\bar B P B)+
{\rm Tr}(\bar B B P)
-\frac{2}{3}{\rm Tr}(\bar B B){\rm Tr}(P)\,, \\
{[\bar B B P]}_S&=&
{\rm Tr}(\bar B B){\rm Tr}(P)\; ,
\end{eqnarray}
where $B$ and $P$ are the baryon octet and pseudoscalar meson nonet
matrices, respectively, given by
\begin{eqnarray}
B&=&\left( \begin{array}{ccc}
\frac{\Sigma^{0}}{\sqrt{2}}+\frac{\Lambda}{\sqrt{6}} & \Sigma^{+} & p \\
\Sigma^{-} & -\frac{\Sigma^{0}}{\sqrt{2}}+\frac{\Lambda}{\sqrt{6}} &n \\
           -\Xi^{-} & \Xi^{0} & -\frac{2\Lambda}{\sqrt{6}}
         \end{array}
\right) \; , \\
P&=&\left(\begin{array}{ccc}
\frac{\pi^{0}}{\sqrt{2}}+\frac{\eta_8}{\sqrt{6}}+\frac{\eta_1}{\sqrt{3}}
& \pi^{+} & K^{+} \\
\pi^{-} & -\frac{\pi^{0}}{\sqrt{2}}+\frac{\eta_8}{\sqrt{6}}
+\frac{\eta_1}{\sqrt{3}}  & K^{0} \\
K^{-} & \bar{K}^{0} & \frac{-2\eta_8}{\sqrt{6}}+\frac{\eta_1}{\sqrt{3}}
                                      \end{array}
              \right) \; .
\end{eqnarray}
The part of the interaction Lagrangian which allows for a fluctuation
of the proton into kaons and hyperons is contained in
\begin{eqnarray}
\mathcal{L}_I&=&-g_1\bar{p}\eta_1p
+g_8\biggl[\bar{p}\pi^0+\frac{1-4\alpha}{\sqrt{3}}\bar{p}\eta_8
+\frac{1+2\alpha}{\sqrt{3}}\bar{\Lambda}K^-
+(2\alpha-1)\bar\Sigma^0K^-
\nonumber\\
&-&\sqrt{2}\bar{n}\pi^-+\sqrt{2}(2\alpha-1)\bar\Sigma^-K^0\biggr]p
 + \cdots
\end{eqnarray}
The final states resulting from pseudoscalar meson emission by the
proton are summarized as
 \begin{eqnarray}
|\Psi \rangle &\sim& -g_1|p\eta_1\rangle
+g_8\biggl[\frac{1-4\alpha}{\sqrt{3}}|p\eta_8\rangle
+|p\pi^0\rangle+\frac{1+2\alpha}{\sqrt{3}}|\Lambda K^+\rangle
\nonumber\\
&+&(2\alpha-1)|\Sigma^0 K^+\rangle
  -\sqrt{2}|n\pi^+\rangle+\sqrt{2}(2\alpha-1)|\Sigma^+K^0\rangle\biggr] \,.
\label{eq:chiralp}
\end{eqnarray}
In the absence of the fluctuation, the proton is made up of the conventional two $u$ quarks and one $d$ quark.
Thus $\Psi(p)$ may be interpreted as the 5-quark component of the proton wave function which is given by
\begin{eqnarray}\label{5q-ChQM}
|uuds\bar{s}\rangle^{\rm ChQM}= G_{1}|\Sigma^0K^+\rangle
+G_{2}|\Sigma^+K^0\rangle+G_{3}|\Lambda^0K^+\rangle+
G_4|p\eta_1\rangle+G_{5}|p\eta_8\rangle+ \,,
\end{eqnarray}
where the $G_i$ are the coefficients corresponding to the respective factor
in Eq.~(\ref{eq:chiralp}).
Each component in the last equation can be represented
in terms of quark cluster comnfigurations as
\begin{eqnarray}\label{K-Y}
|p\eta_{1,8}\rangle=|(uud)_{1/2}(s\bar{s})_0\rangle_{1/2}\,, \quad
|\Sigma^0K^+\rangle =|(uds)_{1/2}(u\bar{s})_0\rangle_{1/2}\,,\nonumber \\
|\Sigma^+K^0\rangle
=|(uus)_{1/2}(d\bar{s})_{0}\rangle_{1/2}\,, \quad |\Lambda^0K^+\rangle =
|(usd)_{1/2}(u\bar{s})_{0}\rangle_{1/2} \,.
\end{eqnarray}

\subsection{Proton wave function including general configurations of the
$uuds$ subsystem}
Another, more general form of the 5-quark component was proposed and
analyzed in
Ref.~\cite{An:2005cj}. Instead of first generating a meson coupling to a baryon
cluster, they consider the genuine 5-quark or $q^4\bar{q}$ pentaquark component
in the proton.
In this model the 5-quark component in this model
may be expressed in terms of the $uuds$ and the $\bar s$ wave functions as
\begin{equation}
|uuds\bar{s}\rangle ^{uuds}= |(uuds)\bar{s}\rangle_{1/2}.
\end{equation}
The flavor wave
functions for the $uuds\bar{s}$ components are usually constructed
by coupling the $uuds$ to the $\bar{s}$ flavor wave function.
The configurations studied in~\cite{An:2005cj} include at most one unit of
orbital angular momentum. The favored configurations are connected
to a positive sign for the strangeness magnetic moment and a negative one
for the strangeness contribution to the proton spin.

\section{Strangeness magnetic moment and spin of the proton}\label{sec:2}

In the nonrelativistic quark model the strangeness magnetic moment
operator $\vec{\mu }_s$ and the strangeness contribution to the
proton spin operator $\vec{\sigma}_s$ are defined as
\begin{equation}
\vec{\mu }_s=\frac{e}{2m_s}\underset{i}{ \sum }\widehat{S}_i(\widehat{\ell}_s +\widehat{\sigma}_s)\; ,
\end{equation}
\begin{equation}
\vec{\sigma}_s=\widehat{\sigma}_s + \widehat{\sigma}_{\bar{s}} \;.
\end{equation}
$\widehat{S}_i$ is the strangeness counting operator with
eigenvalue +1 for an $s $ and -1 for an $ \bar{s}$ quark and $m_s$
is the constituent mass of the strange quark. To calculate the matrix
elements of these operators explicit forms of the
spin-flavor wave functions of the proton including orbital
angular momentum are needed.

For the first model the
spin-flavor wave function can be constructed by coupling the
$|s\bar{s}\rangle_{j_s=0,1}$ configuration to the $|uud\rangle
_{1/2}$ cluster. Since the admixed $s\bar s$ carries negative
intrinsic parity, an orbital P-wave ($\ell=1$) has to be introduced
into the nucleon quark cluster wave function. The simplest configuration
(see also Ref.~~\cite{Henley:1991ge}) corresponds to an 1S-state of
the $s\bar s$ pair moving in a p-wave relative to the (uud) valence
quark cluster of the nucleon. Then the 5-quark component
with total angular momentum 1/2  can be written in the general form:
\begin{equation}\label{5q-sea-spin+1/2}
|uuds\bar{s} \rangle^{s\bar{s}}_{\frac{1}{2},m_{ps\bar{s}}
=\frac{1}{2}}=\underset{j_s,j_i=0,1}{ \sum
}\alpha_{j_sj_i}|[(s\bar{s})_{j_s}\otimes \ell=1]_{j_i }\otimes
(uud)_{\frac{1}{2} }\rangle_{\frac{1}{2},m_{ps\bar{s}}=\frac{1}{2}}
\end{equation}
with the normalization $\underset{j_s,j_i=0,1}{ \sum
}|\alpha_{j_sj_i}|^2=1$.

Similarly, for the proton wave function in the ChQM, where the sea-quark contributions
are embedded in the pseudoscalar mesons, a relative $P$-wave between the pseudoscalars
and  the $uud$ or hyperon clusters has to be included.
The spin-flavor wave function with spin +1/2 for each coupled meson-baryon state of
Eq.~(\ref{K-Y}) may be expressed as
\begin{equation}
|uuds\bar{s}
\rangle^{\rm ChQM}_{\frac{1}{2},\frac{1}{2}}=|[(q\bar{s})_{j_s=0}\otimes
\ell=1]_{j_i }\otimes (qqs)_{s
}\rangle_{\frac{1}{2},m_{ps\bar{s}}=\frac{1}{2}}.
\end{equation}
%%%%%%%%%%%%%%%%%%%%%%%%%%%%%%%%%%%%%%%%%%%%%%%%%%%%%%%%%%%%%%%%%%%%%%%%%%%

Wave functions of the pentaquark $uuds\bar{s}$ states employed in the
third model are more complicated because no restrictions are set
concerning the sub-clusters.
One has to carefully consider the coupling of the color, spin,
flavor and spatial parts to
construct the total wave functions~\cite{An:2005cj}. The color part of the
antiquark in the pentaquark states is a $[11]$ antitriplet, denoted
by the Weyl tableau of the SU(3) group. Hence
the color symmetry of all the $uuds$ configurations is limited to
a $[211]$ triplet in order to form a pentaquark color singlet labeled by the
Weyl tableau $[222]$.
Three flavor symmetry patterns exist for the $uuds$ system corresponding to the octet
representation for the proton: $[31]_F ,[22]_F $ and $[211]_F $
characterized by the $S_4$ Young tableau. However, the pentaquark should
be antisymmetric under any permutation of the four quark
configuration. If the spatial wave function is symmetric, the
spin-flavor part of the $uuds$ component must be a $[31]$ state in
order to form the antisymmetric color-spin-flavor $uuds$ part of the
pentaquark wave function. For instance, the flavor symmetry
representations $[31]_F$ and $[211]_F$ may combine with the spin
symmetry state $[22]_S $ to form the mixed symmetry spin-flavor
states $[31]_{FS}$ (the explicit forms may be found in
\cite{An:2005cj, Bijker:2004gr, An:2008xk}). In this work we
consider only the case that the $uuds$ component is in the ground
state with the spin symmetry $[22]_S$ corresponding to spin zero,
and the relative orbital angular momentum between the $uuds$
component and the $\bar{s}$ is of one unit to obtain the positive
parity for the proton wave function.

%%%%%%%%%%%%%%%%%%%%%%%%%%%%%%%%%%%%%%%%%%%%%%%%%%%%%%%%%%%%%%%%%%%%%%%%%%%

Theoretical results for the strangeness magnetic moment $\mu_s$
of the proton and the strangeness contribution to the proton spin
$\sigma_s$ are listed in Table I. In the first model we have fixed
the configuration parameters as
$\alpha_{1,0}=\alpha_{1,1}=\bar{\alpha}$. The
strangeness magnetic moment $\mu_s$ depends explicitly on
$\alpha_{0,1}$, which is related
to the amplitude for the $s\bar{s}$ quark cluster with spin 0.
Setting $\alpha_{0,1}=0$ is equivalent to
excluding the quantum number $J^{PC}=0^{-+}$ for the $s\bar{s}$
admixture in the nucleon wave function connected to the the production of $\eta$
and $\eta'$ in $N\bar N$ annihilation as discussed in \cite{Dover:1990ic,Ellis:1999er}.
The chiral quark model always gives results for $\mu_s$ and $\sigma_s$
which are negative, the size of the strangeness contribution depends
on the coupling $g_8^2$.
For the third model, we show only the results for the cases where the
$uuds$ component is in the ground state with the spin-flavor
configurations $[31]_{FS}[211]_{F}[22]_{S}$ and
$[31]_{FS}[31]_{F}[22]_{S}$ and the relative motion between the
$uuds$ component and the $\bar{s}$ is a $P$-wave.
\begin{table}
\begin{center}\label{spin-table}
\begin{tabular}{c c c c  }
\hline
$|uuds\bar{s}\rangle$ & $\mu_s(\frac{eB^2}{2m_s})$ & $\sigma_s(B^2)$&
 \\
\hline
&&&\\
$ s\bar{s} $& $-0.55\bar{\alpha}\alpha_{0,1}$ & $-1.22\bar{\alpha}^2$ \cite{Gutsche:1997gy}&  \\
&&&\\
$\rm{ChQM}$& $-1.1g_8^2$ & $-0.31g_8^2$&  \\
&&&\\
$[31]_{FS}[211]_{F}[22]_{S}$& $ -\frac{1}{3}$\cite{An:2005cj}& $-\frac{1}{3}$\cite{An:2005cj}&  \\
&&&\\
$[31]_{FS}[31]_{F}[22]_{S} $& $ -\frac{1}{3}$\cite{An:2005cj}& $-\frac{1}{3}$\cite{An:2005cj}&  \\
&&&\\
\hline
\end{tabular}
\normalsize \caption{Strangeness magnetic moment and spin of the
proton for the three models of the 5-quark component.}
\end{center}
\end{table}

All the three models yield negative values for the strangeness
contribution to the proton spin, which is consistent with present
experimental results \cite{Ashman:1987hv,Ellis:1994py}. Negative
values for the strangeness magnetic moment also result from all three models.
Note that we restricted the considerations of Ref.~\cite{An:2005cj}
to the pentaquark components with the $uuds$ configurations
$[31]_{FS}[211]_{F}[22]_{S}$ and $[31]_{FS}[31]_{F}[22]_{S} $,
respectively.

\section{$N \bar N$ transition amplitude and branching ratios} \label{sec:3}

To describe the annihilation reactions $N\bar N \to X \phi $
$(X=\pi^0 , \eta ,\rho^0, \omega )$ we use an effective transition dynamics,
which is evaluated in the context of a simple constituent quark model.
In this specific process the $\phi $ meson couples to the intrinsic $s\bar s$
component of the nucleon, which is the leading order OZI allowed contribution.
The process $p\bar{p}$ annihilation
into $\phi X$ involving the 5-quark components in the proton wave
function can be described by the quark line diagrams of Fig. 1.
In the hadronic transition the effective quark annihilation operator
is taken with the quantum numbers of the vacuum ($^3P_0$, isospin $I=0$ and
color singlet). Meson decays and $N\bar N$ annihilation into two mesons
are well described phenomenologically using such an effective
quark-antiquark vertex. At least fro meson decay, this approximation has been
given a rigorous basis in strong-coupling QCD. The nonperturbative
$q\bar{q}$ $^3P_0$ vertex is defined according to
\cite{Dover:1992vj}
\begin{equation}
V^{ij}=\sum_\mu\sigma^{ij}_{-\mu}Y_{1\mu}(\vec{q}_i-\vec{q}_{j})\delta^{(3)}(\vec{q}_i+\vec{q}_{j})(-1)^{1+\mu}1^{ij}_F1^{ij}_C~,
\end{equation}
where $Y_{1\mu}(\vec{q})=|\vec{q}|\mathcal{Y}_{1\mu}(\widehat{q})$
with $\mathcal{Y}_{1\mu}(\widehat{q})$ being the spherical harmonics
in momentum space, and $1^{ij}_{F}$ and $1^{ij}_{C}$ are unit
operators in flavor and color spaces, respectively. The spin
operator $\sigma^{ij}_{-\mu}$ is part of the $^3P_0$ vertex,
destroying or creating quark-antiquark pairs with spin 1.

            \begin{figure}[htb]
            \begin{center}
            \includegraphics[width=5cm]{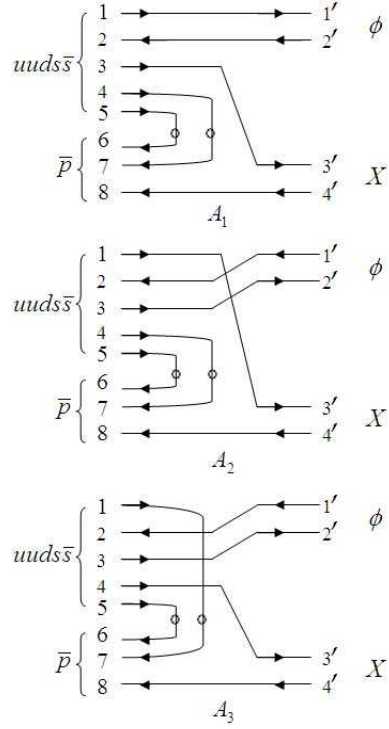}
            \end{center}
\caption{Quark line diagrams for the production of two meson final states in $p\bar{p}$ annihilation.
Small circles refer to the effective vertex of the $^3P_0$ quark dynamics for $q\bar{q}$ annihilation.
The first diagram corresponds to the shake-out of
the intrinsic $s\bar{s}$ component of the proton wave function \cite{Ellis:1994ww,Gutsche:1997gy}.}
            %\label{fig1}
            \end{figure}

In the momentum space representation the transition amplitudes for
the quark diagrams of Fig. 1 are given by
\begin{eqnarray} T_{A_I} = \int d^3q_1
..d^3q_8 d^3q_{1'}..d^3q_{4'}\langle \phi
X|\vec{q}_{1'}..\vec{q}_{4'} \rangle
\langle\vec{q}_{1'}..\vec{q}_{4'}|
\mathcal{O}_{A_I}|\vec{q}_{1}..\vec{q}_{8}\rangle  \langle
\vec{q}_{1}..\vec{q}_{8}|(uuds\bar{s})\otimes(\bar{u} \bar{u}
\bar{d})\rangle
\end{eqnarray}
where $(\bar{u} \bar{u} \bar{d})$ stands for the antiproton wave
function and $(uuds\bar{s})$ for the five quark component of the
proton wave function. The effective operators $\mathcal{O}_{A_I}$ take the
form
\begin{equation}
\mathcal{O}_{A_1} = \lambda_{A_1}
\delta^{(3)}(\vec{q}_1-\vec{q}_{1'})
\delta^{(3)}(\vec{q}_2-\vec{q}_{2'})\delta^{(3)}(\vec{q}_3-\vec{q}_{3'})\delta^{(3)}(\vec{q}_8-\vec{q}_{4'})V^{56}V^{47}~,
\end{equation}
\begin{equation}
\mathcal{O}_{A_2} = \lambda_{A_2} \delta^{(3)}(\vec{q}_2-\vec{q}_{1'}) \delta^{(3)}(\vec{q}_3-\vec{q}_{2'})\delta^{(3)}(\vec{q}_1-\vec{q}_{3'})\delta^{(3)}(\vec{q}_8-\vec{q}_{4'})V^{56}V^{47}~,
\end{equation}
\begin{equation}
\mathcal{O}_{A_3} = \lambda_{A_3} \delta^{(3)}(\vec{q}_2-\vec{q}_{1'}) \delta^{(3)}(\vec{q}_3-\vec{q}_{2'})\delta^{(3)}(\vec{q}_4-\vec{q}_{3'})\delta^{(3)}(\vec{q}_8-\vec{q}_{4'})V^{56}V^{17}~.
\end{equation}
The $\delta$-functions represent the noninteracting and continuous
quark-antiquark lines in the diagrams.  The constants $\lambda_{A_I}$
describe the effective strength of the transition topology and are
considered to be overall fitting parameters in the
phenomenological description of experimental data. Since the 5-quark
component is treated as a small perturbative admixture in the
proton ($B^2<<1$), we ignore the transition amplitude with a term
to $ \langle \vec{q}_{1}..\vec{q}_{8}|(uuds\bar{s})\otimes(\bar{u}
\bar{u} \bar{d}\bar{s}s)\rangle$ or the rearrangement process
\cite{Ellis:1994ww}.

In this work the internal spatial wave functions are taken in the
harmonic oscillator approximation. For the mesons $ M $ ($ \phi$ and
$ X $), the wave function can be expressed in terms of the quark
momenta as
\begin{equation}
\langle M |\vec{q}_{i'}
\vec{q}_{j'}\rangle\equiv\varphi_M(\vec{q}_{i'},\vec{q}_{j'})\chi_{M}=N_M{\rm
exp}\left\{-\frac{R^2_M}{8} \Big(\vec{q}_{i'} -\vec{q}_{j'}\Big)^2
\right\}\chi_{M},
\end{equation}
with $N_M = (R_M^2/\pi)^{3/4}$ and $R_M$ is the meson radial
parameter. The spin-color-flavor wave function is denoted by
$\chi_M$. The baryon wave functions are given by
\begin{equation}
\langle B|\vec{q}_i \vec{q}_j \vec{q}_k\rangle\equiv\varphi_B\chi_{B}=
N_B{\rm exp}\left\{-\frac{R^2_B}{4}
\Big[(\vec{q}_j-\vec{q}_k)^2
+\frac{(\vec{q}_j+\vec{q}_k-2\vec{q}_i)^2}{3}\Big]  \right\}
\chi_{B} \,,
\end{equation}
where $N_B=(3R^2_B/\pi)^{3/2}$ and $R_B$ is the baryon radial
parameter. For the first and the second model the full 5-quark
component wave function, resulting from the coupling of a meson to
a baryon, is given by
\begin{eqnarray}
\langle \vec{q}_1\cdot\vec{q}_5 |uuds\vec{s}\rangle&=&
\varphi_{uuds\bar{s} }(\vec{q}_1,\cdots,\vec{q}_5)\chi_{uuds\bar{s}}
=
N_{uuds\bar{s}} \, {\rm exp}\left\{-\frac{R^2_{B}}{4}
\Big[(\vec{q}_4-\vec{q}_5)^2
+\frac{(\vec{q}_4+\vec{q}_5-2\vec{q}_3)^2}{3}\Big]  \right\}
\nonumber \\
&\times&
{\rm exp}\left\{-\frac{R^2}{8}
(\vec{q}_3+\vec{q}_4+\vec{q}_5-\vec{q}_1-\vec{q}_2)^2
\right\}
Y_{1\mu}(\vec{q}_3+\vec{q}_4+\vec{q}_5-\vec{q}_1-\vec{q}_2)
\nonumber\\
&\times& {\rm exp}\left\{-\frac{R^2_{M}}{8} (\vec{q}_1-\vec{q}_2)^2 \right\}
\ (\chi_{B}\otimes\chi_{M} ).
\end{eqnarray}
The exponential form with the radial parameter $R$ and the spherical
harmonics $Y_{1\mu}$ together represent the internal relative P-wave
between the 3-quark and 2-quark clusters.

For the third model the proton wave function includes a pentaquark
component $uuds\bar{s} $ with the $uuds$ part in the ground state
and the P-wave internal relative orbital angular momentum between
$uuds$ and the $\bar s$. One may write the spatial wave
function of the pentaquark component $uuds\bar{s} $ as
\begin{eqnarray}
\varphi_{uuds\bar{s} }(\vec{q}_1,\cdots,\vec{q}_5)&=&
N_{uuds\bar{s}} \,
{\rm exp}\biggl\{-\frac{R^2_{B}}{4}\Big[(\vec{q}_2-\vec{q}_3)^2
+\frac{(\vec{q}_2+\vec{q}_3-2\vec{q}_4)^2}{3} \nonumber\\
&+&\frac{(\vec{q}_2+\vec{q}_3+\vec{q}_4-3\vec{q}_5)^2}{6}+
\frac{(\vec{q}_2+\vec{q}_3+\vec{q}_4+\vec{q}_5-4\vec{q}_1)^2}{10}\Big]\biggr\}
\nonumber \\
&\times&
Y_{1\mu}\biggl(\frac{\vec{q}_2+\vec{q}_3+\vec{q}_4
+\vec{q}_5-4\vec{q}_1}{\sqrt{20}}\biggr) \; .
\end{eqnarray}

By choosing the plane wave basis for the relative motion of the
proton and antiproton, the initial state wave functions in the
center of momentum system
($\vec{k}=\vec{q}_1+\vec{q}_2+\vec{q}_3+\vec{q}_4+\vec{q}_5$) are
obtained as:
\begin{equation}
\langle \vec{q}_{1}\cdots\vec{q}_{8}|(uuds\bar{s})\otimes
(\bar{u} \bar{u} \bar{d})\rangle=
\varphi_{uuds\bar{s},\bar p}[\chi_{uuds\bar{s}}\otimes\chi_{\bar p}]_{S,S_z}
\end{equation}
with
\begin{equation}
\varphi_{uuds\bar{s},\bar p}=\varphi_{uuds\bar{s}}\varphi_{\bar
p}\delta^{(3)}(\vec{q}_1+\vec{q}_2
+\vec{q}_3+\vec{q}_4+\vec{q}_5-\vec{k})
\delta^{(3)}(\vec{q}_6+\vec{q}_7 +\vec{q}_8+\vec{k})~.
\end{equation}
The spins of the $p\bar p$ system are coupled to the total spin $S$
with projection $S_z$. Similarly, the final state $\phi X $ wave
functions in the center of momentum system are given by
($\vec{q}=\vec{q}_{1'}+\vec{q}_{2'}$):
\begin{equation}
\langle \phi X|\vec{q}_{1'}...\vec{q}_{4'}\rangle=\varphi_{\phi,X}[\chi_{\phi }\otimes\chi_{X}]_{j_i,m_{\epsilon }}
\end{equation}
with
\begin{equation}
\varphi_{\phi,X}=\varphi_{\phi}\varphi_{X}\delta^{(3)}
(\vec{q}-\vec{q}_{1'}-\vec{q}_{2'})
\delta^{(3)}(\vec{q}+
\vec{q}_{3'}+\vec{q}_{4'})~.
\end{equation}
The spins of the two meson states are coupled to $j_i $
with projection $m_{\epsilon } $.

In the low-momentum approximation, the transition amplitude $T_{fi}$
of the annihilation reaction of the $S$-wave $\overline pp$ initial
state $i $ to the $P$-wave two-meson final state $f $ with the quark
line diagrams $A_I$ as shown in Fig. 1 is derived as
\begin{equation}\label{T-1}
T_{fi}(\vec{q},\vec{k})=\lambda_{A_I}F_{L=0,\ell_f=1}q \,
{\rm exp} \left\{ -Q^2_q q^2 -Q^2_k k^2\right\} \langle f | O_{A_I}|i \rangle
\end{equation}
The index $i $ represents the initial state $^{2I+1,2S+1} L_J$ where
$L $ is the orbital angular momentum, $S$ is the total spin, $J $ is
the total angular momentum and $I$ is the total isospin. The final
state $f $ is represented by the set of quantum numbers $f=\{ \ell_f j
J'\}$ where $\ell_f $ is the relative orbital angular momentum. The
constants $F_{0,1}$, $Q^2_q$ and $Q^2_k$ are geometrical constants
depending on the radial parameters. The matrix element $\langle f |
O_{A_I}|i \rangle$ is the spin-flavor weight for a quark line
diagram $A_I$. The detailed evaluation of the expression in
Eq.(\ref{T-1}) is given in Appendix A. Since the in the particle
basis $p\bar p$ and $n\bar n$
give the same spin-flavor weight, the $\phi$ production from the
nucleon-antinucleon annihilation at rest can be described by the
transition amplitude Eq.(\ref{T-1}) multiplied with a factor
$\sqrt{2}$.

As we consider $p\bar{p}$ annihilations at rest where the strong
interaction between the proton and antiproton may largely distort
the $\overline pp$ hydrogen-like wave function at small distances
\cite{Yan:1997yi}, the effect of the initial state interaction is in
general not negligible. The inclusion of the initial state interaction
for the atomic state of the $p\bar{p}$ system results in the
transition amplitude \cite{Kercek:1999sc},
\begin{equation}\label{atomic state}
T_{f,LSJ}(\vec{q})=\int d^3k ~ T_{fi}(\vec{q},\vec{k}) \phi ^I_{LSJ}(\vec{k} ),
\end{equation}
where $\phi ^I_{LSJ}(\vec{k} ) $ is the protonium wave function in
momentum space for fixed isospin $I $. The partial decay width for
the transition of the $p\bar{p} $ state to the two-meson state $\phi
X $ is given by
\begin{equation}
\Gamma_{p\bar{p}\rightarrow \phi X}=\int\frac{d^3p_\phi}{2E_\phi}\frac{d^3p_X}{2E_X}\delta ^{(3)}(\vec{p}_\phi +\vec{p}_X)\delta(E-E_\phi-E_X )|T_{f,LSJ}(\vec{q})|^2
\end{equation}
where $E$ is the total energy ($E=1.876$ GeV) and $E_{\phi ,X}=
\sqrt{m^2_{\phi ,X}+\vec{p}^2_{\phi ,X}} $ is the energy of outgoing
meson $\phi$ and $X$ with mass $m_{\phi ,X}$ and momentum
$\vec{p}_{\phi ,X}$. With the explicit form of the transition
amplitude given by Eq.~(\ref{T-1}), the partial decay width
for the S to P transition ($L=0 $, $ \ell_f=1 $) is written as
\begin{equation}\label{decay width}
\Gamma_{p\bar{p}\rightarrow \phi X}=\lambda_{A_I}^2f(\phi ,X)\langle f | O_{A_I}|i \rangle ^2 \gamma(I,J) ,
\end{equation}
with
\begin{equation}
\gamma(I,J)= |F_{0,1}  \int d^3 k ~\phi ^I_{LSJ}(\vec{k} ) { \rm exp} \left\{ -Z^2_\gamma k^2\right\}|^2
\end{equation}
and the kinematical phase-space factor defined by
\begin{equation}\label{phase-space factor}
f(\phi ,X)=2\pi\frac{E_\phi E_X}{E }q^3 {\rm exp} \left\{ -2Z^2_\alpha q^2 \right\}.
\end{equation}
The spin-flavor weights $\langle f | O_{A_I}|i \rangle$ for the transitions $N\bar N \to \phi X$
involving the different 5-quark components of the proton wave functions are listed in Table II.
%%%%%%%%%%%%%%%%%%%%%%%%%%%%%%%%%%%%%%%%%%%%%%%%%%%%%%%%%%%%%%%%%%%%%%%%%%%%%%%%%%%%%%%%%%%%%%%%%%%%%%%%%%%%%%%%%%%%%%%%%%%%%%%%%%%%%%%%
\begin{table}\label{SFtable}
\caption{Spin-flavor matrix elements $\langle f | O_{A_I}|i \rangle$
for the transitions $ p\bar{p}(L=0)\rightarrow\phi X(\ell_f=1)$
which are described by the quark line diagram $A_I$. Here $\eta_{ud}$
refers to the nonstrange flavor combination
$\eta_{ud}=(u\bar{u}+d\bar{d})/\sqrt{2}$.}
\begin{center}
\begin{tabular}{c c c c c c}
\hline
Transition &  $s\bar{s}_{A_1}$ & ChQM
&$[31][31][22]_{A_1}$ &  $[31][211][22]_{A_1}$
 \\
\hline
&&&\\
$^{11}S_0$$\rightarrow\omega\phi$& $\frac{5}{9\sqrt{6}}$& -0.097& $\frac{5}{36\sqrt{6}}$& $\frac{5}{36\sqrt{6}}$\\
&&&\\
$^{33}S_1$$\rightarrow\pi^0\phi$& $\frac{5}{27\sqrt{2}}$& 0.031&$\frac{5}{108\sqrt{2}}$& $\frac{5}{108\sqrt{2}}$\\
&&&\\
$^{31}S_0$$\rightarrow\rho^0\phi$& $\frac{13}{27\sqrt{6}}$&0.040& $\frac{13}{108\sqrt{6}}$& $\frac{13}{108\sqrt{6}}$\\
&&&\\
$^{13}S_1$$\rightarrow\eta_{ud}\phi$& $\frac{1}{9\sqrt{2}}$& 0.013& $\frac{1}{36\sqrt{2}}$& $\frac{1}{36\sqrt{2}}$\\
&&&\\
\hline
\end{tabular}
\end{center}
\end{table}
For the initial values of the total angular momentum $J $ the statistical
weights 1/4 and 3/4 have to be added for $J=0 $ and $J=1 $,
respectively. Finally the branching ratio of S-wave $p\bar{p}$
annihilation to the final state $ \phi X $ is then given by
\begin{equation}\label{BR}
BR(\phi,X)=\frac{(2J+1 )\Gamma_{p\bar{p}\rightarrow \phi X}}{4\Gamma_{tot}(J)},
\end{equation}
where $\Gamma_{tot}(J) $ is the total annihilation width of the
$p\bar{p} $ atomic state with fixed principal quantum number
\cite{Dover:1991mu}.

The model dependence in Eq.(\ref{decay width}) may be reduced by
choosing a simplified phenomenological approach that has been
applied in studies of two-meson branching ratios in
nucleon-antinucleon \cite{Kercek:1999sc} and radiative protonium
annihilation \cite{Gutsche:1998fc}. Namely, instead of the phase
space factor in Eq.(\ref{phase-space factor}) which depends on the
relative momentum and the masses of $\phi X$ system, we use a
kinematical phase-space factor of the form
\begin{equation}\label{f-function}
f(\phi,X)=q\cdot {\rm exp}\{-a_s\,(s-s_{\phi X})^{1/2}\}
\end{equation}
where $a_s=1.2$ GeV$^{-1}$, $s_{\phi X}=(m_{\phi}+m_{X})^{1/2}$ and
$\sqrt{s}=(m_{\phi}^2+q^2)^{1/2}+(m_{X}^2+q^2)^{1/2} $. Last form
is obtained from the fit to the momentum dependence of the cross section of various
annihilation channels \cite{Vandermeulen:1988hh}. In addition, the
functions $\gamma(I,J) $, depending on the initial-state
interaction, are related to the probability for a protonium state to
have isospin $I$ and spin $J $ with the normalization condition
$\gamma(0,J)+\gamma(1,J)=1 $. Here we adopt for a protonium state
the probability $\gamma(I,J) $ and the total decay width
$\Gamma_{tot}(J) $ obtained in an optical potential calculation
\cite{Carbonell:1989cs}, where explicit values are listed
in \cite{Dover:1991mu}.

In Table III we give the theoretical results for the branching ratios
of Eq.~(\ref{BR}) compared with experimental data. The branching
ratios $BR^{s\bar{s}}$, resulting from the first model where the
proton wave function has an explicit $s\bar s$ admixture, have already been
derived and studied in Ref.~\cite{Gutsche:1997gy} by using the same approach.
Annihilation processes in the first and third model are described by the
quark line diagram $A_1$. Since the effective strength parameter
$\lambda_{A_1}$ is a priori unknown it has to be adjusted to data.
For this purpose one entry (as indicated by $\star $) is normalized to the
observed value.

For the second chiral model where the proton wave function contains a
kaon-hyperon or eta-proton cluster component, all three quark line
diagrams may have contributions to the $\overline pp$ annihilation
process. However, the process proceeding by the diagram $A_1$ with the
$|p\eta\rangle$ component in the proton wave function has no
contribution to the transition because of orthogonality to the $\phi $ state.
Therefore, the annihilation process in the second model can only be
described by the quark line diagrams $A_2$ and $A_3$. Considering
the same annihilation pattern in these two diagrams, for
simplicity the two unknown strength parameters
are of the same order with $\lambda_{A_2}= \lambda_{A_3}$. Model
predictions are also normalized to experimental data (as indicated
by $\star $). For final states with $X=\eta $, the physical $\eta $
meson is produced by its nonstrange component $\eta_{ud} $ with
$\eta = \eta_{ud}(\sqrt{1/3}\cos\theta - \sqrt{2/3}\sin\theta) $
corresponding to a variation of the pseudoscalar mixing angle $\theta $
from $\theta=-10.7^o $ to $\theta =-20^o$.

\begin{table}\label{BRtable}
\caption{Branching ratio $BR(\times 10^{4})$ for the transition
$p\bar{p}\rightarrow \phi X$ ($X =\pi^0,\eta,\rho^0,\omega$) in
$p\bar{p}$ annihilation at rest. The results indicated by $\star$
are normalized to the experimental values.}
\begin{center}
\begin{tabular}{c c c c c c}
\hline
Transition & BR$^{\rm exp}$ & BR$^{s\bar{s}}$
& BR$^{\rm ChQM}$ &  BR$^{[31][31][22]}$ & BR$^{[31][211][22]}$\\
\hline
&&&\\
$^{11}S_0$$\rightarrow\omega\phi$ & 6.3$\pm$2.3  & 6.3 $\star$ & 6.3 $\star$ & 6.3 $\star$ & 6.3 $\star$\\
&&&\\
$^{33}S_1$$\rightarrow\pi^0\phi$& 5.5 $\pm$ 0.7& 5.4 & 1.6 & 5.4 & 5.4\\
&&&\\
$^{31}S_0$$\rightarrow\rho^0\phi$& 3.4 $\pm$ 1.0 & 3.8 & 0.87  & 3.8 & 3.8\\
&&&\\
$^{13}S_1$$\rightarrow\eta\phi$& 0.9 $\pm$ 0.3& 1.4$-$1.8 & 0.20$-$0.27 & 1.4$-$1.8 & 1.4$-$1.8\\

&&&\\
\hline
\end{tabular}
\end{center}
\end{table}
As shown in Table III, the theoretical results of the first and
third models, where the proton wave function possesses respectively
a small kaon-hyperon component and a pentaquark, are in good
agreement with the experimental data. Note that for these two cases
the annihilation processes $p\bar{p}\rightarrow \phi X$ are
described with the quark line diagram $A_1$.

\section{Summary}

Three models have been studied for the proton involving intrinsic
strangeness in the form of a 5-quark component $qqqs\bar{s}$ in the
wave function. In particular, the proton wave function is made up of
a $uud$ configuration and a $uud$ cluster with a $s\bar{s}$
sea-quark component, kaon-hyperon clusters based on the simple
chiral quark model, or a pentaquark component $uuds\bar s$. We have
calculated the strangeness magnetic moment $\mu_s $ and spin
$\sigma_s $ for the first and second models and generate negative
values in line with recent experimental indication. Similarly, for
the third model we pick these configurations, where negative values
for $\mu_s$ and $\sigma_s$ result~\cite{An:2005cj}.

We further applied quark line diagrams supplemented by the $^3P_0$ vertex
to study the annihilation reactions $p \bar{p}\rightarrow \phi X$
($X=\pi^0,\eta,\rho^0,\omega$) with the three types of proton wave
functions.  Excellent agreements of the model predictions in the
first and third models with the experimental data are found for the
branching ratios of the reactions of the $L=0$ atomic $p \bar{p}$
state to $\phi X$ ($X=\pi^0,\eta,\rho^0,\omega$).

{\bf Acknowledgements} {\\ \small This work was supported by the DFG
under Contract No. FA67/31-2. This research is also part of the
European Community-Research Infrastructure Integrating Activity
``Study of Strongly Interacting Matter'' (acronym HadronPhysics2,
Grant Agreement No. 227431) and part of the Federal Targeted Program
``Scientific and scientific-pedagogical personnel of innovative
Russia'' Contract No. 02.740.11.0238. We also acknowledge the
generous help of Chun-Sheng AN for providing us with the proton wave
function with the 5-quark component in $uuds$ subsystem used in this
paper. The stay in T$\rm \ddot{u}$bingen of Sorakrai Srisuphaphon
was supported by the DAAD under PKZ:A/07/98879, and the study at SUT
was supported by Burapha University. }
%%%%%%%%%%%%%%%%%%%%%%%%%%%%%%%%%%%%%%%%%%%%%%%%%%%%%%%%%%%%%%%%%%%%%%%%%%%

\numberwithin{equation}{section}
\begin{appendix}
\section{Transition amplitudes of the annihilation processes $p \bar{p}\rightarrow \phi X$}
To describe the annihilation process $p \bar{p}\rightarrow \phi X$
where $X=\pi^0,\eta,\rho^0,\omega$ with the proton wave function
with $s\bar{s}$ sea quark we consider the shake-out of the intrinsic
$s\bar{s}$ component of the proton wave function as indicated in the
diagram $A_1$. With the operator $\mathcal{O}_{A_1} $ and the full
account of the spin-flavor-color-orbital structure of the initial
and final states, the transition amplitude can be written as
\begin{eqnarray}{\label{T-ss}}
T_{if}^{s\bar{s}} =\lambda_{A_1}\langle f| \sum_{\nu,\lambda} (-1)^{\nu+\lambda} \sigma^{56}_{-\nu}\sigma^{47}_{-\lambda}1^{56}_F1^{47}_F1^{56}_C1^{47}_C I_{spatial}^{s\bar{s}} |i\rangle \; ,
\end{eqnarray}
where
\begin{eqnarray}\label{state-i}
|i\rangle=|\{ \chi_{\frac{1}{2},m_{ps\bar{s }}}(uuds\bar{s })\otimes
\chi_{\frac{1}{2},m_{\bar{p }}}
(\bar{u}\bar{u}\bar{d})\}_{S,S_z}\otimes (L,M)\rangle_{J,J_z},
\end{eqnarray}
\begin{eqnarray}\label{state-f}
|f\rangle=| \{\chi_{1,m_\alpha}(\phi)\otimes
\chi_{j_m,m_{3',4'}}(X)\}_{j,m_\epsilon }\otimes
(\ell_f,m_f)\rangle_{J,J_z}.
\end{eqnarray}
The spin-flavor-color content of the clusters is denoted by $\chi
\equiv \chi_{\sigma}\otimes\chi_F\otimes\chi_C$. The 5-quark
component $ \chi_{\frac{1}{2},m_{ps\bar{s }}} (uuds\bar{s })$ is
defined as
\begin{equation}
 \chi_{\frac{1}{2},m_{ps\bar{s }}} (uuds\bar{s })=|
 \{\chi_{j_s,m_s}(s\bar{s})\otimes (\ell=1,\mu)\}_{j_i,m_i }\otimes \chi_{\frac{1}{2},m_p }(uud) \rangle_{\frac{1}{2},m_{ps\bar{s }}}\;.
\end{equation}
The spatial amplitude $I_{spatial}^{s\bar{s }}$ is explicitly given
by
\begin{equation}
I_{spatial}^{s\bar{s }}=\int d^3q_1 ...d^3q_8 d^3q_{1'}...d^3q_{4'} \varphi_{\phi,X }\mathcal{O}_{A_1}^{spatial } \varphi_{uuds\bar{s },\bar{p}}
\end{equation}
where
\begin{eqnarray}
\mathcal{O}_{A_1}^{spatial } =Y_{1\lambda}(\vec{q}_4-\vec{q}_{7})\delta^{(3)}(\vec{q}_4+\vec{q}_{7})Y_{1\nu}(\vec{q}_5-\vec{q}_{6})\delta^{(3)}(\vec{q}_5+\vec{q}_{6})~~~~~
 \nonumber \\ \delta^{(3)}(\vec{q}_1-\vec{q}_{1'}) \delta^{(3)}(\vec{q}_2-\vec{q}_{2'})\delta^{(3)}(\vec{q}_3-\vec{q}_{3'})\delta^{(3)}(\vec{q}_8-\vec{q}_{4'}).
\end{eqnarray}
Partial wave amplitudes can be obtained by projecting the transition
amplitude onto the partial waves, where $L=0$ and $l_f=1$
corresponds to $\overline pp$ annihilation at rest. In the
low-momentum approximation the integrals can be done analytically,
and the partial wave amplitude in the leading order of the external
momenta $q$ is given by
\begin{eqnarray}\label{I-ss}
I_{spatial,L=0,l_f=1}^{s\bar{s}}=qF_{0,1}^{s\bar{s}}f^{s\bar{s}}_{0,1}(\nu,\lambda,\mu,m_f){\exp}\left\{
-Q^2_q q^2 -Q^2_k k^2\right\} \;.
\end{eqnarray}
The geometrical constant $F_{0,1}^{s\bar{s}} $ and the spin-angular momentum function $f_{0,1}^{s\bar{s}}(\nu,\lambda,\mu,m_f) $ are given by
\begin{eqnarray}\label{I-ss1}
F_{0,1}^{s\bar{s}}=2 N \pi ^2 \left(\frac{1}{Q_{p_2}^2}\right)^{3/2}
\left(\frac{3
   \sqrt{\pi }}{\left(Q_{p_4}^2\right){}^{5/2}}-\frac{3 \sqrt{\pi }}{4
   \left(Q_{p_3}^2\right){}^{5/2}}\right),
\nonumber \\ f_{0,1}^{s\bar{s}}(\nu,\lambda,\mu,m_f)=(-1)^{\nu}\delta_{\nu,-\lambda}\delta_{\mu,m_f},~~~~~~~~~~~~
\end{eqnarray}
where $N=N_\phi N_X N_{uuds\bar{s} } N_{\bar{p} }$, and the
coefficients in the exponential expression depend on the meson and
baryon size parameters:
\begin{eqnarray}
Q_k^2&=&\frac{4 R_M^2 R_B^2+9 R^2 R_B^2+3 R_M^2
   R^2}{24 \left(R_M^2+3 R_B^2\right)},
\nonumber\\Q_q^2&=&\frac{12 R_B^4+5 R_M^2 R_B^2+36 R^2 R_B^2+12 R_M^2
   R^2}{24 \left(R_M^2+3 R_B^2\right)},
   \nonumber\\Q_{p_2}^2&=&R_M^2,~Q_{p_3}^2=\frac{1}{2} \left(R_M^2+3 R_B^2\right),
   ~Q_{p_4}^2=2 R_B^2.
\end{eqnarray}

By using the spatial wave amplitude $I_{spatial}^{s\bar{s }}$ we
obtain the transition amplitude $T_{if}^{s\bar{s}} $ taking the form
as in Eq.~(\ref{T-1}) with the spin-color-flavor weight:
\begin{equation}
 \langle f | O_{A_1}|i \rangle=\langle f| \sum_{\nu,\lambda} (-1)^{\nu+\lambda} \sigma^{56}_{-\nu}\sigma^{47}_{-\lambda}1^{56}_F1^{47}_F 1^{56}_C1^{47}_C (-1)^{\nu}\delta_{\nu,-\lambda}\delta_{\mu,m_f}|i\rangle.
\end{equation}
According to the $^3P_0$ quark model the matrix element $ \langle f
| O_{A_1}|i \rangle$ can be evaluated by using the two-body matrix
elements for spin, flavor and color given by
\begin{equation}\label{3p0-spin}
\langle 0 |\sigma^{ij }_\upsilon | \chi^{{J_{ij }}}_{m_{ij }}(ij)  \rangle=\delta_{J_{ij },1}\delta_{m_{ij },-\upsilon}(-1)^\upsilon\sqrt{2},
\end{equation}
\begin{equation}\label{3p0-flavor}
\langle 0 |1^{ij }_F | \chi^{{T_{ij }}}_{t_{ij }}(ij)  \rangle=\delta_{T_{ij },0}\delta_{t_{ij },0}\sqrt{2},
\end{equation}
and
\begin{equation}\label{3p0-color}
\langle0|1^{ij}_{C}|q_{\alpha}^i\bar{q}_\beta^j\rangle=\delta_{\alpha\beta},
\end{equation}
where $\alpha$ and $\beta$ are the color indices. The
spin-color-flavor weights $ \langle f | O_{A_1}|i \rangle$ are
evaluated for various transitions, as listed in Table II.

In case of the simple chiral quark model the annihilation processes are
described by the quark line diagrams $A_2$ and $A_3$.
Then the transition amplitude is set up as
\begin{equation}\label{T-ChQM-1}
T^{ChQM}_{if}=T^{\rm ChQM}_{if}(\mathcal{O}_{A_2})
+T^{\rm ChQM}_{if}(\mathcal{O}_{A_3}),
\end{equation}
where the corresponding transition amplitudes for the two quark line
diagrams are given by
\begin{eqnarray}\label{T-A2}
T^{\rm ChQM}_{if}(\mathcal{O}_{A_2})
= \lambda_{A_2}\langle f| \sum_{\nu,\lambda} (-1)^{\nu+\lambda}
\sigma^{56}_{-\nu}\sigma^{47}_{-\lambda}1^{56}_F1^{47}_F
1^{56}_C1^{47}_C I_{spatial,A_2}^{\rm ChQM} |i\rangle
\end{eqnarray}
and
\begin{eqnarray}\label{T-A3}
T^{\rm ChQM}_{if}(\mathcal{O}_{A_3})
=\lambda_{A_3}\langle f| \sum_{\nu,\lambda} (-1)^{\nu+\lambda}
\sigma^{56}_{-\nu}\sigma^{17}_{-\lambda}1^{56}_F1^{17}_F 1^{56}_C1^{17}_C
I_{spatial,A_3}^{\rm ChQM} |i\rangle.
\end{eqnarray}
The initial state $|i\rangle$ and the final state $|f\rangle$ take
the same form as defined in Eq.~(\ref{state-i}) and
Eq.~(\ref{state-f}), but the 5-quark component in this case is given
by
\begin{equation}\label{5q-ChQM-chi}
\chi_{\frac{1}{2},m_{KY}}(uuds\bar{s })=\sum^3_{i=1}G_i|
\{\chi^i_{j_s,m_s}(q\bar{s})\otimes (\ell=1,\mu)\}_{j_i,m_i }\otimes
\chi^i_{\frac{1}{2},m_Y}(qqs) \rangle_{\frac{1}{2},m_{KY}},
\end{equation}
where $i=1,2,3$ represent the kaon-hyperon clusters $K^+\Sigma^0$,
$K^0\Sigma^+$ and $K^+\Lambda^0$, respectively, and the coefficients
$G_i$ are as defined in Eq.~(\ref{5q-ChQM}).

In the low-momentum approximation the partial wave amplitude from
each of the quark line diagrams $A1$ and $A2$ in leading order
of the external momentum $q$ takes the general form as in
Eq.~(\ref{I-ss}) but with different coefficients. In order to combine
the two transition amplitudes, we choose the radial parameters for
the baryons and mesons as $R_B=3.1 ~GeV^{-1}$, $R_M=4.1~ GeV^{-1}$
\cite{Gutsche:1997gy} and the size parameter between the two quark
clusters as $R=4.1~ GeV^{-1}$.  Then the total transition
amplitude eq.(\ref{T-ChQM-1}) becomes
\begin{equation}\label{T-ChQM-A2+A3}
T^{\rm ChQM}_{if}=\lambda_{\rm ChQM}F^{\rm ChQM}_{0,1}q~ {\rm exp}
\left\{ -Z^2_q q^2 -Z^2_k k^2\right\} \langle f | O_{\rm ChQM}|i \rangle,
\end{equation}
where $F^{\rm ChQM}_{0,1}=4.9\times10^{-4}~ GeV^{-11}$, $Z_q\simeq2.3~
GeV^{-1}$ and $Z_k\simeq1.3~ GeV^{-1}$, and
$\lambda_{A_2}=\lambda_{A_3}=\lambda_{\rm ChQM}$. The total
spin-color-flavor weight $ \langle f | O_{\rm ChQM}|i \rangle$ is
calculated with the spin-angular momentum wave functions in Eq.
(\ref{5q-ChQM-chi}) and its elements are derived as
\begin{equation}
f_{0,1}^{\rm ChQM}=-(-1)^\nu\delta_{\nu,-\lambda}\delta_{\mu,m_f}+2
(-1)^\mu\delta_{\mu,-\nu}\delta_{\lambda,m_f}
+2(-1)^\lambda\delta_{\mu,-\lambda}\delta_{\nu,m_f}~.
\end{equation}

Finally we discuss the third model where the proton wave function
includes a $5q$ component in the form of a pentaquark configuration.
The $\phi$
production is described by only the quark line diagram $A_1$, and
the transition amplitude takes the same form as eq.(\ref{T-ss}) but
the 5-quark component $|uuds\bar{s }\rangle $ is given by
\begin{equation}
\chi_{\frac{1}{2},m_{ps\bar{s }}} (uuds\bar{s })=|
\{\chi_{1/2,m_{\bar{s}}}(\bar{s})\otimes (\ell=1,\mu)\}_{j_i,m_i
}\otimes \chi_{s,s_z}(uuds) \rangle_{\frac{1}{2},m_{ps\bar{s }}}.
\end{equation}
In the low-momentum approximation, the partial wave amplitude and
for the transition of the $S$-wave $\overline pp$ state to the
$P$-wave two-meson final states takes the same form as
Eq.~(\ref{I-ss}). The spin-angular momentum function
$f_{0,1}^{s\bar{s}}(\nu,\lambda,\mu,m_f)$ is also the same as the
one in Eq.~(\ref{I-ss1}) but the corresponding geometrical constant
is given by
\begin{eqnarray}
F_{0,1}=-\frac{3}{16}  \sqrt{5} N \pi ^4
   \left(\frac{1}{Q_{p_2}^2}\right)^{3/2}
   \left(\frac{\left(\frac{1}{Q_{p_4}^2}\right)^{3/2}}{\left(Q_{p_3}^
   2\right)^{5/2}}-\frac{4
   \left(\frac{1}{Q_{p_3}^2}\right)^{3/2}}{\left(Q_{p_4}^2\right)^{
   5/2}}\right),
\end{eqnarray}
with the constants depending on the baryon and meson size
parameters:
\begin{eqnarray}
 &\;&Q_k^2=\frac{7 R_B^2}{30}-\frac{R_B^4}{2 \left(3 R_B^2+R_M^2\right)},
~Q_q^2=\frac{1}{8} R_B^2 \left(5-\frac{R_B^2}{3 R_B^2+R_M^2}\right),
\nonumber\\  &\;& Q_{p_2}^2=R_B^2+\frac{R_M^2}{2},
~Q_{p_3}^2=\frac{1}{2} \left(3 R_B^2+R_M^2\right),
~Q_{p_4}^2=2 R_B^2.
  \end{eqnarray}

\end{appendix}

\end{document}